\documentclass[reprint,aps,prl,twocolumn,superscriptaddress,floatfix,unsortedaddress]{revtex4-1}
\usepackage{graphicx}
\usepackage{float}
\usepackage{epstopdf}
\usepackage{natbib}
\usepackage{hyperref}
\usepackage{color}
\usepackage{amsmath}
\usepackage{amssymb}
\usepackage{ulem}

\begin{document}

\title{Laser Accelerated Ions from a Shock Compressed Gas Foil}

\author{M. H. Helle}
\altaffiliation{U.S. Naval Research Laboratory, 4555 Overlook Ave., SW, Washington, DC 20375}
\email[Corresponding author: ]{mike.helle@nrl.navy.mil}
\author{D. F. Gordon}
\author{D. Kaganovich}
\author{Y. Chen}
\altaffiliation{Research Support Instruments, 4325-B Forbes Boulevard, Lanham, MD 20706}
\author{J. P. Palastro}
\author{A. Ting}

\date{\today}

\begin{abstract}
We present results of energetic laser-ion acceleration from a tailored, near solid density gas target. Colliding hydrodynamic shocks compress a pure hydrogen gas jet into a 70 $\mu$m thick target prior to the arrival of the ultra-intense laser pulse. A density scan reveals the transition from a regime characterized by a wide angle, low energy beam to one of a more focused beam with a high energy halo. In the latter case, three dimensional simulations show the formation of a Z-pinch driven by the axial current resulting from laser wakefield accelerated electrons. Ions at the rear of the target are then accelerated by a combination of space charge fields from accelerated electrons and Coulombic repulsion as the pinch dissipates.    

\end{abstract}


\maketitle

High-energy ions have been accelerated by means of laser interactions with over-dense or tenuous plasmas for over a decade\cite{macchi2013ion,albright2007relativistic}. These investigations relied on varying accelerating mechanisms, the most prominent being Laser Hole Boring, Target Normal Sheath Acceleration (TNSA)\cite{fuchs2006laser}, Laser Blowout Acceleration (BOA), and Radiation Pressure Acceleration (RPA). All of these mechanisms primarily rely on accelerating fields produced by charge separation between the massive ions and laser accelerated hot electrons.  While progress continues to be made, these experiments have relied on solid targets that can limit repetition rate and purity of the accelerated ions. 

Recently long-wave infrared laser acceleration of ions by means of electro-static shock acceleration in gas targets has become a reality\cite{palmer2011monoenergetic,haberberger2012collisionless}. These experiments take advantage of short pulse CO$_{2}$ lasers ($\lambda_{0}$ =10.6 $\mu$m) that have a critical plasma density within the operational range of gas jets (10$^{18}$-10$^{20}$ cm$^{-3}$), where the critical density (n$_{crit}$ = (2$\pi$c)$^{2}$ m$_{e}\epsilon_{0}$/e$^{2} \lambda_{0}^{2}$) is the cutoff density at which electromagnetic waves no longer propagate within the plasma. 

While this type of acceleration technique may be difficult to achieve for ultrashort pulse near-infrared lasers since it requires strong electron heating, the advantages of gas targets remain; namely they are relatively simple, produce high purity targets with variable densities, and can be operated at high repetition rates. Unfortunately, current gas jets produce densities well below the critical density for near-infrared wavelengths, where most of the world's terawatt and petawatt lasers operate.

\begin{figure}
\includegraphics[width= 3.0 in]{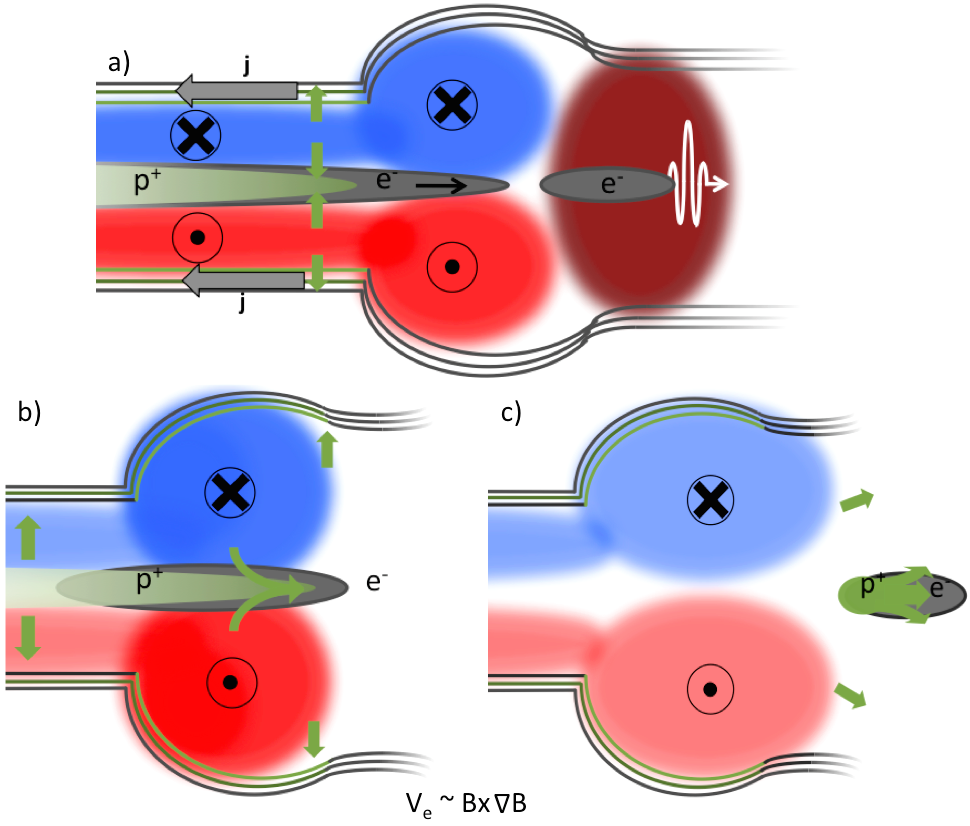}
\caption{\label{MechanismIllustration} Illustration of the acceleration process. a) An intense laser pulse drives a self-modulated wakefield, injecting and accelerating a high energy electron beam from the ambient plasma. A Z-pinch of the trailing electrons and ions is driven by the high current beam.  b) The high magnetic field and field gradients lead to a axial guiding center drift for a secondary electron beam moving out of the plasma. c) The space charge fields of the exiting electrons accelerate ambient plasma ions. The Z-pinched compressed ions explode out radially, while cold ions collectively escape with the electron beam.   }
\end{figure}

In this letter, we report the acceleration of protons form a new type of target, a ``gas foil'', that has been developed at the Naval Research Laboratory\cite{gordon2013laser,kaganovich2014shaping}. The target is based on the propagation of strong hydrodynamic shocks in gas. The shock front acts to increase the local density to values near critical density for near-infrared wavelengths ($\sim$10$^{21}$ cm$^{-3}$) and steepen the density gradients. Gradients at both the front and rear of the target are important in that they reduce deleterious laser propagation effects, including ionization defocusing and filamentation, while at the same aiding the production of energetic particles as they are accelerated out of the rear of the target. Additionally, the target can be operated at various densities and at high repetition rates. In previous work\cite{kaganovich2010measurements}, it was observed that the gas flow returns to its pre-shocked state in $\sim$$\mu$s. Thus repetition rates in excess of 10 kHz are possible.  Here, we will present results of $\sim$2MeV proton acceleration from this target using only 500mJ of laser energy as well as a discussion of the relevant mechanisms, including TNSA and Magnetic Vortex 
Acceleration. 

Magnetic Vortex Acceleration is a recently proposed and as of yet experimentally confirmed acceleration technique that exhibits a myriad of physical processes\cite{bulanov2000generation,willingale2008longitudinal,nakamura2010high,sentoku2000high,bulanov2010generation,gordon2013laser}. In particular, this scheme relies on intense magnetic field gradients produced at a plasma gradient to drive acceleration, fig. \ref{MechanismIllustration}. Using a gas foil target to access this mechanism provides a route to generate high-energy ions at high repetition from a high purity source that leverages compact solid-state lasers.

\begin{figure}
\includegraphics[width= 3 in]{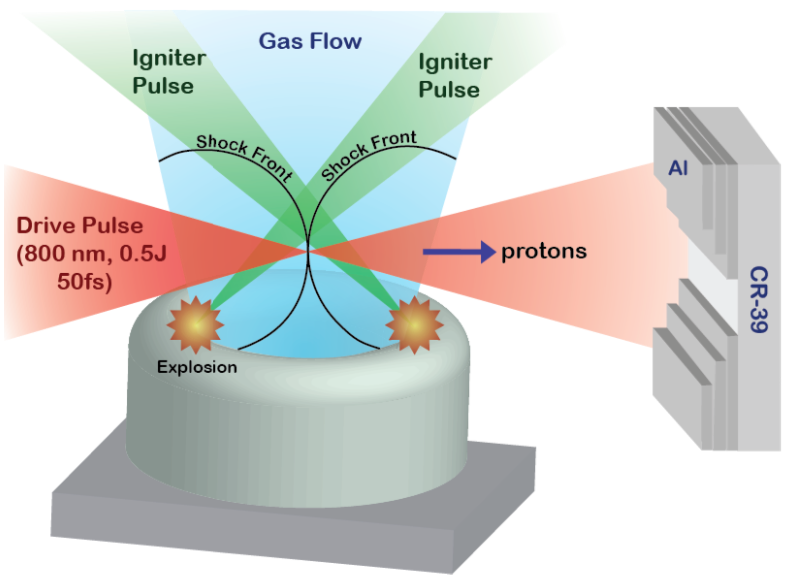}
\caption{\label{Setup} A schematic of the experimental setup including the CR-39 detector stack. A 1mm supersonic gas jet delivers the neutral gas while laser ignited strong hydrodynamic shocks act to shape it. The accelerated ions are then detected by CR-39 plates with varying Al filters.}
\end{figure}

The experimental set-up is shown in fig. \ref{Setup}.  Strong hydrodynamic shockwaves are ignited in a standard gas jet in vacuum using nanosecond frequency doubled Nd:YAG pulses. The individual shocks produce a profile that consists of a sharp leading edge, followed by a slow falling edge, with a peak value that is ($\gamma$+1)/($\gamma$-1), where $\gamma$ is the ratio of specific heat. This value can range from 4 to 10 dependent on the degree of excitation of the internal degrees of freedom of the gas. When two shocks are allowed to collide, the resulting peak density is the sum of the individual shock densities. This technique  provides the density increase necessary to approach critical density. For these acceleration experiments two blastwaves were used to tailor the hydrogen gas flow into a 75 $\mu$m thick gas ``foil'' with 30 $\mu$m gradients and tunable peak densities from 2.5 - 5$\times$10$^{20}$ molecules cm$^{-3}$. This provides a pure hydrogen target at densities and gradients previously unexplored. The formation of the target is monitored using a 400nm, 50fs laser probe that goes to shadowgraphic and interferometric diagnostics [Supplemental].

The interaction is driven by a 500 mJ, 800 nm, 50 fs laser pulse generated by the TFL laser system at NRL. The pulse is focused, using an f/2 off-axis parabola (OAP), to a vacuum spot size of 2.6 $\mu$m, 1/e$^{2}$, reaching a peak intensity of 1$\times$10$^{20}$ W cm$^{-2}$. We focused the beam at the front edge of the ``foil'' to allow the beam to relativistically self-focus within the density ramp. This allows the beam to reach higher intensities. The energy and spatial distribution of the ions is measured using a 1 mm thick CR-39 plate with aluminum filters of varying thickness spaced perpendicular to the laser polarization axis. The stack was placed 13 cm from the gas jet to prevent laser damage of the plate.

\begin{figure}
\includegraphics[width= 3.5 in, height= 2 in]{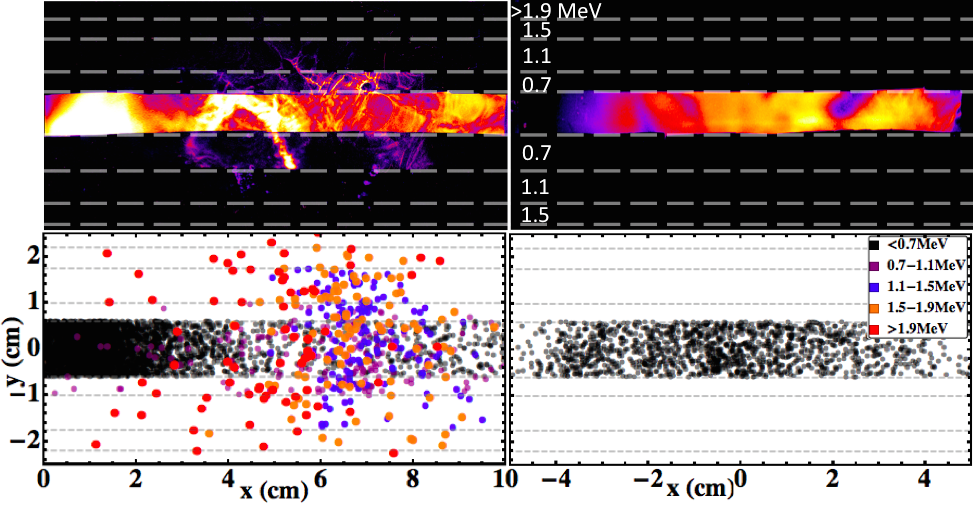}
\caption{\label{ExperimentalDistribution}  False color image of a 10cm x 5cm CR-39 plates for the cases of 0.3 n$_{crit}$ (left) and 0.6 n$_{crit}$ (right). The dashed lines represent the location of the filters and their lower energy bound is listed on the left. On the bottom are plots of simulated test particles projected out to the plane of the detector color-coded to the experimental energy buckets.}
\end{figure}

A density scan was performed over the operational range of our gas jet at intervals of 0.5$\times$10$^{20}$ molecules cm$^{-3}$. Assuming fully ionized hydrogen, this corresponds to peak plasma densities of n$_{e}$ = 0.3-0.6 n$_{crit}$ It was observed that at densities above 0.4 n$_{crit}$, a proton beam contained within the length of detector was accelerated in the forward direction. The beam was unable to penetrate even a single filter layer (9 $\mu$m) and thus had peak energy of $<$700 keV. Upon closer examination of the tracks produced, the pit diameters were $<$10$\mu$m corresponding to the proton energies $<$200 keV \cite{seguin2003spectrometry}. An example detector image is given in fig. \ref{ExperimentalDistribution}. 

At a density of 0.3 n$_{crit}$, we observed protons penetrating through the 9$\mu$m, 18$\mu$m, and 27$\mu$m thick filters with no proton penetrating the 36$\mu$m thick filter (fig. \ref{ExperimentalDistribution}). This indicates a broad energy distribution of protons with a maximum energy between 1.5-1.9 MeV. Unlike the higher density cases, the spatial distribution of the protons within the unshielded region extends beyond the length of the detector. It is characterized by an intense beam near the left boundary with a highly nonuniform distribution to the right of that. These uniformities are consistent with the distribution of higher energy protons observed in the shielded regions above and below.

\begin{figure}
\includegraphics[height= 2.1 in]{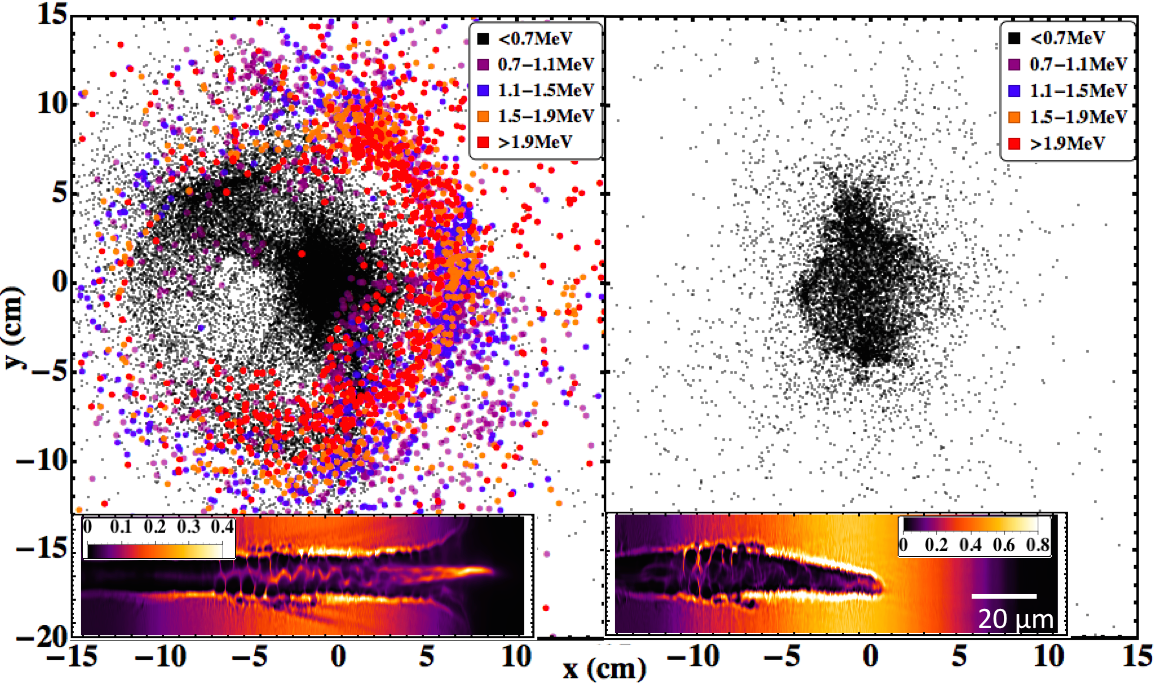}
\caption{\label{Breakout} Plots of the full test particles spatial distribution as projected out to the plane of the detector for 0.3n$_{crit}$ (left) and 0.6 n$_{crit}$ (right). Inset are proton density plots taken along the polarization plane for each case. Note that for the 0.3n$_{crit}$ case the cavitation region extends entire length of the plasma while it only extend halfway through for the 0.6 n$_{crit}$.}
\end{figure}

The origin of these accelerated protons was investigated using the TurboWAVE 3D particle-in-cell code. The simulation was initialized using the experimental parameters and density profiles. Test particles were placed within the simulation box, their orbits were tracked, and projected to the detector plane. The projected ion test particles for the cases of 0.3n$_{crit}$ and 0.6n$_{crit}$ are given in figs. \ref{ExperimentalDistribution} and \ref{Breakout}. In fig. \ref{ExperimentalDistribution} the test particle spatial distributions are filtered to match the experimental conditions and show excellent agreement. To understand the underlying process, the full simulated distribution with the corresponding proton density profiles are given in fig. \ref{Breakout}. For the 0.6n$_{crit}$ case, the protons are contained within a beam width of 8 cm FWHM and have a peak energy of 140 keV, consistent with the experimental track diameters. Examining the results of the PIC simulation, we see that the laser pulse is unable to penetrate the target, fig. \ref{Breakout}. The protons that are accelerated in the forward direction originate from the rear side of the target. They are accelerated by space-charge fields produced from hot electrons generated by the laser-target interaction exiting the rear of the target. This is consistent with the Target Normal Sheath Acceleration mechanism (TNSA)\cite{fuchs2006laser}. 

\begin{figure}
\includegraphics[width= 3.5 in]{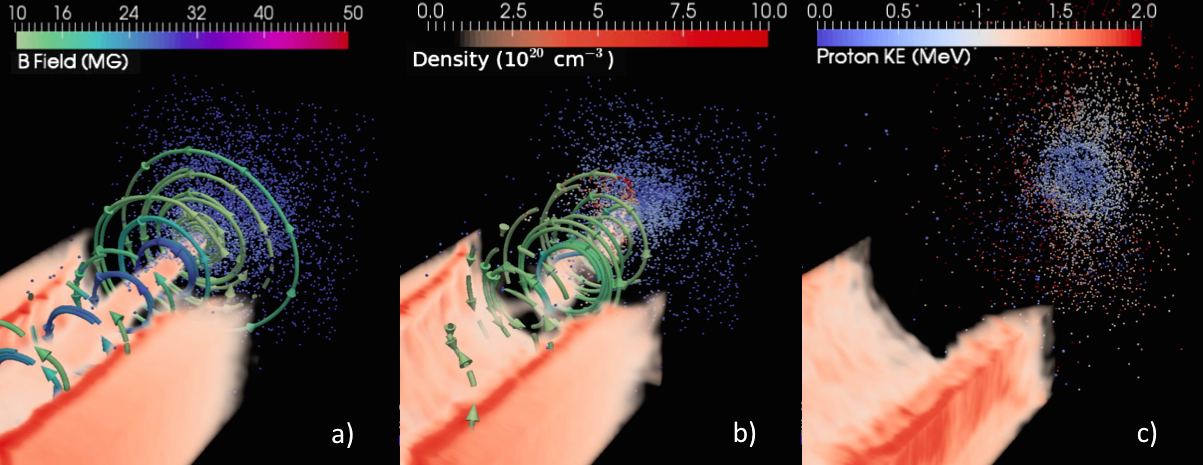}
\caption{\label{VolumetricRendering} 3D volumetric renderings of the electron density and azimuthal fields with proton test particles included. The test particles are color coded by their kinetic energy. Note that test particles have a radial energy distribution with lowest energy on-axis and the highest off-axis. A movie of these simulation results are provided as supplemental material.}
\end{figure}

Examining the 0.3n$_{crit}$ case, we see a vastly different situation. The forward accelerated protons are characterized by a low energy beam on-axis with a halo of high-energy protons with energies $\lesssim$2MeV. When examining the proton density plot, we observe that at these densities the laser pulse is able to penetrate through the entire plasma. The Z-pinch is also clearly seen extending through the plasma. To better illustrate the dynamic process that is occurring, 3D renderings showing the evolution of the electron plasma density, magnetic field lines, and proton test particles are given in fig. \ref{VolumetricRendering}. Initially, the intense laser pulse undergoes self-focusing and modulation instability in the density up ramp of the plasma region. The pondermotive force of the pulse drives electron cavitations. Plasma electrons become trapped and accelerated within these cavitations. Due to the slow plasma wave phase velocity and short dephasing length, electrons are easily injected, accelerated, and then stream out of the trapping region of the wakefield. These fast electrons trail behind the laser front, forming an axial fast current. To maintain plasma quasi-neutrality, a cold electron return current is formed to balance the fast current. The two oppositely signed currents repel one another, forming an axial fast current and a cylindrical-shell cold return current. The on-axis fast current has an average current of ~22kA. This process leads to the generation of a large ($\sim$50MG) azimuthal magnetic field contained within the region. During this process, the mobile ions pinch on-axis and an electron-ion pinch is formed with n$_{e}$$\approx$n$_{i}$$\approx$.3n$_{crit}$ and radius of $\sim$1$\mu$m. This whole process trails the intense laser pulse until it exits the plasma region.

At the exit, the laser pulse ponderomotively expels the ambient electrons from the density ramp. After the intense laser pulse exits the plasma, the fast electron current and large azimuthal magnetic fields begin to flow out, and a situation reminiscent of the Magnetic Vortex acceleration mechanism occurs. The protons at the interface undergo the same inward pinch as those inside the main plasma region, however the magnitude of pinch rapidly drops off due to the drop in density. The fast current eventually leaves the plasma. At this point the magnetic fields begin to dissipate and the protons, no longer confined, explode outward radially. At the same time they acquire longitudinal momentum from the space-charge fields setup by the escaping fast current. These protons are those observed in the high-energy halo. Additionally, there exist a low-density population of protons outside the ramp. These are accelerated collectively by the fast electrons leaving the plasma region. These protons are the source of low energy protons on axis. Examining the test particle energy in the final rendering, it can be seen that the proton beam exhibits an energy distribution that is radially dependent as in the experiments. 

\begin{figure}
\includegraphics[width= 3 in]{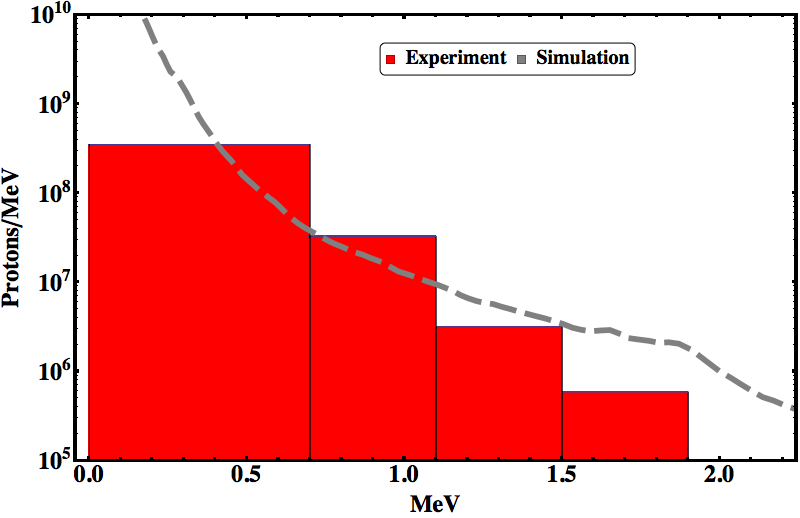}
\caption{\label{Energy Plot} Plot comparing experimental to simulated proton energy spectrum. Proton distribution asymmetries,  simulated experimental parameters, or sampling statistics with PIC macroparticles may be the cause of discrepancies at higher energies.}
\end{figure}

Using the test particle results, we directly compared the proton energy spectrum to that observed in the experiments. The results are plotted in fig. \ref{Energy Plot}. Pit counting of the CR-39 plate was preformed by producing a composite image from a motorized scanning microscope. The pits were then counted using the cell counting algorithm within the image analysis software, ImageJ. In the analysis of the experimental proton energy spectrum, we assumed that the proton distributions were symmetric about the beam axis. Additionally, we subtracted out higher energy totals from from lower energy bands to prevent double counting. The results show general agreement between experiment and simulations. The simulations do predict higher energy protons, however this maybe due to proton distribution asymmetries, discrepancies in the simulated experimental conditions, or sampling statistics with PIC macroparticles. 


The overall agreement between the experiments, simulations, and theory aids in validating our results. Scaling these results to existing simulation work, we observed much lower ion energies\cite{bulanov2000generation,willingale2008longitudinal,nakamura2010high,sentoku2000high,bulanov2010generation,gordon2013laser}. This disagreement is directly due to the highly 3D nature of the process. Namely, reducing the process to one transverse dimension constrains the degrees of freedom afforded to the pinch. This was observed when 2D simulations with identical initial conditions to the ones above were preformed\cite{helle2015laser}. While predicting a lower peak density, these simulations produced a peak proton energy of 13MeV, much higher than what was observed in experiment and 3D simulations. The momentum that would normally go into the second transverse dimension is instead distributed between the other two dimensions resulting in a higher overall kinetic energy. It's interesting to note that the 2D simulations were able to produce the same energy for the on-axis beam. This is due to the longitudinal accelerating force produced by the forward directed electrons. 

The presented acceleration mechanism is directly controlled by the laser and plasma parameters. By tailoring the plasma peak density, thickness, and gradients it could be possible to explore regions of higher magnetic fields and produce even higher energy beams. The target provides a potential route to high repetition operation as new, rep-rated laser systems are developed.

This work was supported by the Department of Energy and the Naval Research Laboratory Base Program. We would like to acknowledge helpful discussions with B. Hafizi, S. Bulanov, J. Penano, and A. Zingale. 


\end{document}


\section{Supplemental}

\begin{figure}[h]
\includegraphics[width= 3 in]{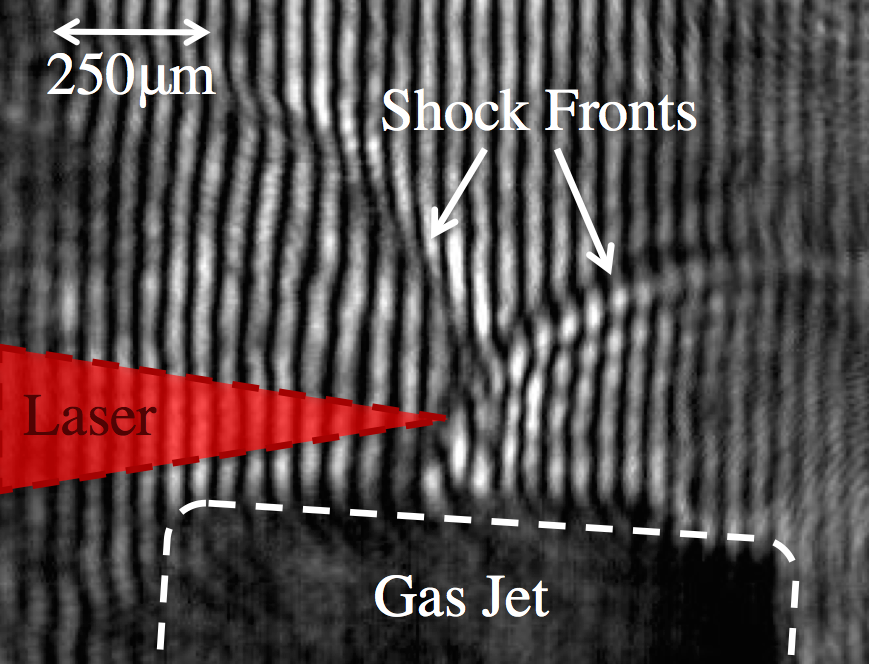}
\caption{\label{Interferogram} An interferogram of two colliding shocks in a supersonic gas flow. The intense drive laser is focused at the front of the colliding shock region. The shocks form a high density, 70 $\mu$m thick, pure hydrogen target for ion acceleration.  }
\end{figure}

To actively monitor and measure the density profile of the target, a small portion of the USPL beam was frequency doubled and used as a probe. The probe was split in two with one part going to a shadowgraphy setup and the other used in a folded wave interferometer. An example interferogram of colliding shocks is given in fig. \ref{Interferogram}. The target is sufficiently thick and dense that when ionized that it is impossible to extract the phase information from the interferograms. However, this is not the case for the neutral density interferograms and, under the assumptions that the gas behaves as an ideal gas, we are able to extract the density profiles using an Inverse Able transform.The extracted density profile is corresponding to the above interferogram is shown in fig. \ref{DensityProfile}.

\begin{figure}[hb]
\includegraphics[width= 3 in]{DensityProfile}
\caption{\label{DensityProfile} The corresponding neutral density profile extracted from fig. \ref{Interferogram} using an Inverse Able transform.}
\end{figure}